\documentclass[twocolumn,journal]{IEEEtran}
\usepackage[T1]{fontenc}
\usepackage[utf8]{inputenc}
\usepackage{units}
\usepackage{amsmath}
\usepackage{amsthm}
\usepackage{amssymb}
\usepackage{graphicx}
\usepackage[unicode=true,
 bookmarks=true,bookmarksnumbered=true,bookmarksopen=true,bookmarksopenlevel=1,
 breaklinks=false,pdfborder={0 0 0},pdfborderstyle={},backref=false,colorlinks=false]
 {hyperref}
\hypersetup{pdftitle={Your Title},
 pdfauthor={Your Name},
 pdfpagelayout=OneColumn, pdfnewwindow=true, pdfstartview=XYZ, plainpages=false}
\usepackage{breakurl}

\makeatletter
\let\oldforeign@language\foreign@language
\DeclareRobustCommand{\foreign@language}[1]{%
  \lowercase{\oldforeign@language{#1}}}
\theoremstyle{plain}
\newtheorem{thm}{\protect\theoremname}
\theoremstyle{definition}
\newtheorem{defn}[thm]{\protect\definitionname}
\theoremstyle{definition}
\newtheorem{example}[thm]{\protect\examplename}
\theoremstyle{remark}
\newtheorem{rem}[thm]{\protect\remarkname}

\usepackage[caption=false,font=footnotesize]{subfig}
\usepackage{todonotes}

\@ifundefined{showcaptionsetup}{}{%
 \PassOptionsToPackage{caption=false}{subfig}}
\usepackage{subfig}
\makeatother

\providecommand{\definitionname}{Definition}
\providecommand{\examplename}{Example}
\providecommand{\remarkname}{Remark}
\providecommand{\theoremname}{Theorem}

\begin{document}

\title{Towards quantification of incompleteness in the pairwise comparisons
method}

\author{Konrad~Kułakowski,~\IEEEmembership{IEEE Member,} Anna Prusak,
Jacek Szybowski \thanks{Konrad Kułakowski, (corresponding author), AGH University of Science
and Technology, the Department of Applied Computer Science, Poland,
e-mail: \protect\href{mailto:konrad.kulakowski@agh.edu.pl}{konrad.kulakowski@agh.edu.pl}.}\thanks{Anna Prusak, Cracow University of Economics, the Department of Quality
Management, Poland, e-mail: \protect\href{mailto:anna.prusak@uek.krakow.pl}{anna.prusak@uek.krakow.pl}.}\thanks{Jacek Szybowski, AGH University of Science and Technology, the Faculty
of Applied Mathematics, Poland, e-mail: \protect\href{mailto:szybowsk@agh.edu.pl}{szybowsk@agh.edu.pl}.}}

\markboth{preprint}{Your Name \MakeLowercase{\emph{et al.}}: Your Title}
\maketitle
\begin{abstract}
Alongside consistency, completeness of information is one of the key
factors influencing data quality. The objective of this paper is to
define ways of treating missing entries in pairwise comparisons (PC)
method with respect to inconsistency and sensitivity. Two important
factors related to the incompleteness of PC matrices have been identified,
namely the number of missing pairwise comparisons and their arrangements.
Accordingly, four incompleteness indices have been developed, simple
to calculate, each of them take into account both: the total number
of missing data and their distribution in the PC matrix. A numerical
study of the properties of these indices has been also conducted using
a series of Montecarlo experiments. It demonstrated that both incompleteness
and inconsistency of data equally contribute to the sensitivity of
the PC matrix. Although incompleteness is only just one of the factors
influencing sensitivity, a relative simplicity of the proposed indices
may help decision makers to quickly estimate the impact of missing
comparisons on the quality of final result.
\end{abstract}


\section{Introduction\label{sec:Introduction}}

\subsection{On comparing alternatives in pairs }

\IEEEPARstart{T}{}he pairwise comparisons method is referred to as
a process of comparing objects in pairs to judge which of them is
preferred \cite{Pan2014arpb}. In the PC method, the elements in a
given set are ranked on a pair-by-pair basis (two at a time), until
performing all of the variations. The first evidence of pairwise judgments
comes from the XIII-century philosopher \emph{Ramon Llull} in the
context of the election systems and the social choice theory. This
system was based on binary comparisons \cite{Colomer2011rlfa}. Specifically,
each voting round provides sets of two candidates who should be compared
in pairs, and the winner is the one who gathers a majority of voices
in the highest number of pairwise comparisons. The PC method proposed
by \emph{Llull} was then reinvented and improved by many other scientists
including the XVIII-century French mathematician and philosopher \emph{Nicolas
de Condorcet} \cite{Condorcet1785eota}. In his election system, the
winner (so-called \emph{the Condorcet winner}) is the one who is always
victorious when being compared with any other candidate. However,
\emph{Condorcet} proved that there might be a situation when the winner
cannot exist. He provided a three-voters example (the Condorcet triplet)
when $A$ is preferred over $B$, $B$ over $C$, and $C$ over $A$,
so finally there is no winner {[}Saari 2009{]}. The Condorcet method
was used in the preference aggregation methods of \emph{C. Dodgson}
in 1876 \cite{Klamler2003acot} and \emph{A. H. Copeland} in 1951,
the latter being known as the Copeland’s rule, in which the candidates
are ordered by the number of pairwise wins minus the number of pairwise
defeats \cite{Saari1996tcm,Faliszewski2009lacv}.

 Another scholar known for his contribution to the PC methodology
is an American psychologist and pioneer in psychometric research,
\emph{Louis L. Thurstone}. In 1927 he used \emph{Gaussian distribution}
to analyze pairwise comparisons. His model (also referred to as the
law of comparative judgments) was based on three assumptions: 1) whenever
a pair of stimuli is presented to a respondent it elicits a continuous
preference for each stimulus (which is discriminal process); 2) the
stimulus with higher value in the comparison is preferred by the respondent;
3) these unobserved preferences are normally distributed \cite{MaydeuOlivares2003otmf}.
He linked his approach with the psychophysical theory proposed by
the XIX-century scholars \emph{E. Weber} and \emph{G. Fechner}. The
\emph{Thurstonian model} was reinvented in 1987 by \emph{Y. Takane},
who added a random error to each paired comparison (so-called Thurstone-Takane
model) \cite{MaydeuOlivares2003otmf}. In 1952 \emph{R. A. Bradley}
and \emph{M. E. Terry} proposed an alternate model to the \emph{Thurstonian}
one. They defined the probability that object $j$ ($O_{j}$) is preferred
to object $k$ ($O_{k}$) in a given comparison $c_{jk}$. In the
psychometric approach, the \emph{Bradley-Terry model} is often called
the BTL model, due to its relation to the choice axiom proposed in
1959 by \emph{R. D. Luce} \cite{Dittrich2009flbt,Tsukida2011htap,Vista2016mpcu}. 

The widely known application of PC method is the \emph{Analytic Hierarchy
Process} (AHP) and the \emph{Analytic Network Process} (ANP), the
multi-criteria decision support techniques developed in the 1970s
by the American mathematician, \emph{T. L. Saaty} \cite{Saaty2008rmai}.
Besides the AHP/ANP methods, other multi-criteria decision techniques
based on comparisons of alternatives include ELECTRE, PROMETHEE or
MACBETH \cite{Greco2005mcda}. However, only the AHP/ANP judgments
result in real numbers representing the relative strength of preference
\cite{Kulakowski2016note}.

Despite its long history, the PC method (especially with relation
to the AHP/ANP and the pairwise comparison matrices) is among the
prevalent topics in recent studies, exploring problems such as inconsistency
\cite{Cernanova2018iosc}, rank reversal \cite{Mufazzal2018anmc,Wang2006aata}
and incomplete judgments \cite{Pan2014arpb}. These characteristics
are important indicators of data quality, which plays a critical role
in modern decision-making processes, especially at business and governmental
level \cite{Batini2009mfdq}. 

\subsection{Quality of data}

The literature does not provide a universal set of data quality dimensions.
Discrepancies in types and definitions of the quality characteristics
are due to the contextual nature of data quality. According to \emph{Batini}
et al. \cite{Batini2009mfdq}, the six most essential classifications
of data quality criteria have been provided by \emph{Wand and Wang}
\cite{Wand1996adqd}, \emph{Wang and Strong} \cite{Wang1996bawd},
\emph{Redman} \cite{Redman1997dqft}, \emph{Jarke} et al. \cite{Jarke2003fodw},
\emph{Bovee} et al. \cite{Bovee2003acfa}, and \emph{Naumann} \cite{Naumann2002qdqa}.
The analysis of these classifications allowed to distinguish a set
of four attributes of data quality most commonly described in the
literature. They include accuracy, completeness, consistency, and
timeliness. 

The first of them, accuracy, is “the extent to which the data is correct,
reliable and certified” \cite{Wang1996bawd}. \emph{Redman} \cite{Redman1997dqft}
defines this term as a measure of the proximity of a data value ($v$)
to other values ($v'$). \emph{Batini} et al. \cite{Batini2009mfdq}
distinguish two types of accuracy, namely syntactic and semantic,
specifying that data quality methodologies only consider syntactic
accuracy, indicating the closeness of $v$ to the corresponding definition
domain $D$. In DAMA report \cite{DAMA2003tspd} data accuracy is
defined as “the degree to which data correctly describes the ‘real
world’ object or event being described,” and its measure is “the degree
to which the data mirrors the characteristics of the real world object
or objects it represents.” 

Completeness was also defined in multiple ways, for example, “the
ability of an information system to represent every meaningful state
of a real-world system” \cite{Wand1996adqd}, or “percentage of real-world
information entered in data sources and/or data warehouse” \cite{Jarke2003fodw}.
The authors of \cite{DAMA2003tspd} defined this criterion as “the
proportion of stored data against the potential of \textbf{$100\%$}
complete” measured as “the absence of blank (null or empty string)
values or the presence of non-blank values.” In the literature, completeness
is often associated with missing values, which exist in the real world
but not in the database \cite{Batini2009mfdq}. This criterion is
crucial in pairwise comparison context when some pairs of objects
remain with no comparisons, so only partial information is available.
This causes other issues such as problems with calculating inconsistency
of a partially filled matrix \cite{Bozoki2010ooco}. More information
on this criterion is given in  Section \ref{sec:Montecarlo-section}
of this paper. 

Consistency is one of the fundamental characteristics of data quality
but defined in many different ways. \emph{Blake and Mangiamelli} \cite{Blake2009etsa}
emphasized that consistency is a multidimensional concept that can
be represented by three aspects: representational consistency, integrity,
and semantic consistency. Representational consistency refers to the
presentation of data in the same format and compatibility with other
(e.g., previous) data. Data integrity requires fulfilling four constraints:
entity, referential, domain and column. Importantly, violations of
entity integrity may lead to redundant or incomplete data. Semantic
consistency indicates no contradiction between different data values
in a particular set. In the literature, the most commonly discussed
are representative consistency in relation to databases and semantic
consistency in relation to the introduced values. Concerning the PC
method, consistency of data is often regarded in terms of inconsistency
indices \cite{Brunelli2013apoi,Kulakowski2018iito}. The mathematical
basis of this measures is provided in Section \ref{sec:Indices-of-incompleteness}. 

The last but not least is timeliness. Interpretation of this attribute
is different across the literature. Thus, \emph{Batini} et al. \cite{Batini2009mfdq}
suggested it should be considered in a broader context, as the time-related
dimension. According to \emph{Wand and Wang} \cite{Wand1996adqd},
timeliness refers to “the delay between a change of a real-world state
and the resulting modification of the information system state.” Other
time-related dimensions are currency, interpreted as “the degree to
which a datum is up-to-date” \cite{Redman1997dqft} or “when the information
was entered in the sources and/or the data warehouse” \cite{Jarke2003fodw}. 

\subsection{Motivation and the organization of the manuscript}

Many scientific articles deal with the inconsistency in the pairwise
comparisons method. Thus, also many methods for measuring inconsistency
have been proposed and thoroughly investigated. As a guide in this
rich literature may serve the works \cite{Koczkodaj2014oaoi,Brunelli2013iifp,Brunelli2013apoi}.
Amazingly the same does not apply to incompleteness. Although some
researchers have proposed methods for calculating the ranking for
incomplete paired comparisons, the influence of incompleteness to
the final result has not been sufficiently studied. One of the exceptions
here can be Harker \cite{Harker1987ipci}, but even this work does
not provide us with the methods to measure incompleteness. Therefore,
the purpose of this research is to determine the impact of the incompleteness
of data on the correctness of the ranking in the PC method. During
the work, we have identified two critical factors related to the incompleteness
affecting the quality of data. These are the number of missing pairwise
comparisons and the arrangement of missing comparisons. For this reason,
we propose four incompleteness indices, where each of them depends
on both: the total number of missing data and their distribution in
the pairwise comparisons matrix. The performed Montecarlo experiments
confirmed their usefulness as fast and quick tests of data quality.

The presented paper is composed of \ref{sec:Summary} sections including
introduction (Section \ref{sec:Introduction}) and summary (Section
\ref{sec:Summary}). Section \ref{sec:Preliminaries} outlines the
theory of the pairwise comparisons method and the PC matrices, explaining
phenomena such as incompleteness, inconsistency, and sensitivity.
Section \ref{sec:Indices-of-incompleteness} presents four groups
of incompleteness indices ($\alpha\text{-index}$, $\beta\text{-index}$,
tree index, and the compound index) allowing for determining to what
extent a given PC matrix based ranking is at risk due to the incompleteness
of data. In Section \ref{sec:Montecarlo-section}, numerical experiments
are presented demonstrating relationship between incompleteness, inconsistency
and sensitivity. 

\section{Preliminaries\label{sec:Preliminaries}}

\subsection{Pairwise comparisons}

\begin{figure}[tbh]
\begin{centering}
\includegraphics[scale=0.5]{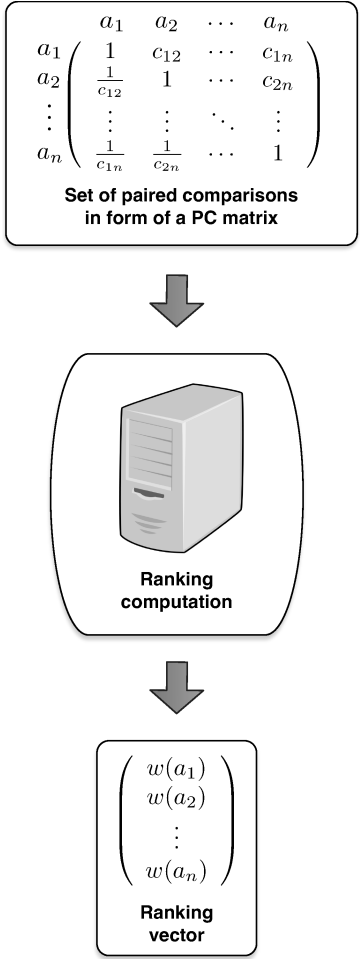}
\par\end{centering}
\caption{From alternatives to ranking - the pairwise comparisons approach}

\label{fig:pairwise-comparisons-scheme}
\end{figure}

The pairwise comparisons method very often is used as a way that allows
experts to create a ranking based on a series of individual comparisons.
The subjects of comparisons are alternatives. Beginning the ranking
procedure experts compare alternatives in pairs. Then the results
of individual comparisons are used as an input to the appropriate
mathematical procedure, which allows computing the final numerical
ranking (Fig. \ref{fig:pairwise-comparisons-scheme}). 

Let $A=\{a_{1},\ldots,a_{n}\}$ be a finite set of alternatives representing
options among which a decision maker can choose. Similarly, let $C=\{c_{ij}\in\mathbb{R}_{+}:i,j=1,\ldots,n\}$
be a set of expert judgements about each pair $(a_{i},a_{j})\in A\times A$,
so that $c_{ij}$ is the result of comparisons $a_{i}$ against $a_{j}$.
Assigning the certain real value $v\in\mathbb{R}_{+}$ represents
the expert's opinion that the alternative $a_{i}$ is $v$ times more
important than $a_{j}$. It is convenient to represent the set of
comparisons in the form of a matrix $C=(c_{ij})$, hereinafter referred
to as the PC (pairwise comparisons) matrix. Since a comparison of
a given alternative to itself does not indicate the advantage of any
of the two alternatives being compared, the diagonal of $C$ is composed
of ones. Similarly, in most of the cases, it is assumed that if $a_{i}$
is $v$ times more important than $a_{j}$ than also $a_{j}$ is $v$
times less important then $a_{i}$. The latter observation leads to
the equality $c_{ij}=1/c_{ji}$. In such a case it is convenient to
use the following definition. 
\begin{defn}
A matrix $C=(c_{ij})$ is said to be reciprocal if for all $i,j=1,\ldots,n$
holds $c_{ij}=1/c_{ji}$.
\end{defn}
The pairwise comparisons method aims to transform the set of paired
comparisons (i.e., the PC matrix) into the ranking vector (Fig. \ref{fig:pairwise-comparisons-scheme}).
Let us define the function that assigns the weight (also called as
the importance or the priority) to every single alternative. Every
\emph{PC matrix} can also be naturally presented in the form of a
graph. 
\begin{defn}
Let $G_{C}=(V,E,L)$ be a labelled, directed graph with the set of
vertices $V=\{a_{1},\ldots,a_{n}\}$, the set of edges $E\subseteq V\times V\backslash\{(a_{1},a_{1}),\ldots,(a_{n},a_{n})\}$,
and the labelling function $L:E\rightarrow\{c_{1,2},\ldots,c_{n,n-1}\}$
so that $L(a_{i},a_{j})=c_{ij}$. $G_{C}$ is said to be induced by
the matrix $C$. 
\end{defn}
In such a graph vertices correspond to alternatives and edges correspond
to the comparisons among the alternatives. 
\begin{defn}
\label{def:Let-the-output}Let the output degree of $a_{i}$ be denoted
by $\textit{outdeg}(a_{i})$ and be given as 
\[
\textit{outdeg}(a_{i})=\left|\{j:(a_{i},a_{j})\in E\}\right|
\]
\end{defn}
It is easy to observe that the output degree of vertex $a_{i}$ is
equal to the number of comparisons of alternative $a_{i}$ with others. 
\begin{defn}
The ranking function for $A$ is a function $w:A\rightarrow\mathbb{R}_{+}$
that assigns a positive real number to every alternative $a\in A$. 
\end{defn}
The role of the ranking computation procedure is to determine the
value of $w$ concerning every alternative. The list of all values
$w(a_{1}),\ldots,w(a_{n})$ we will often write in the form of a transposed
vector $w$: 
\begin{equation}
w=[w(a_{1}),\ldots,w(a_{n})]^{T},\label{eq:ranking_vector}
\end{equation}

Very often $w$ is called interchangeably as a priority or weight
vector. There are several methods of transforming paired comparisons
into the ranking. According to the most popular one, referred to in
the literature as eigenvalue method (EVM), the ranking is formed as
the appropriately rescaled principal eigenvector \cite{Saaty1977asmf}.
Thus, to calculate $w$ in EVM one have to solve equation 
\begin{equation}
Cw_{\textit{max}}=\lambda_{\textit{max}}w_{\textit{max}},\label{eq:eigenvector-equation}
\end{equation}

where $\lambda_{max}$ is the spectral radius (principal eigenvalue)
of $C$, then rescale $w$ so that all its entries sum up to $1$. 

\[
w=[s\cdot w_{\textit{max}}(a_{1}),\ldots,s\cdot w_{\textit{max}}(a_{n})]^{T},
\]

where 
\[
s=\left(\sum_{i=1}^{n}w_{\textit{max}}(a_{i})\right)^{-1}.
\]

There are a dozen other weighting methods for PC matrices \cite{Jablonsky2015aosp,Wang2007peit,Yuen2012mmpm,Kou2014acmm,Dong2008acso}.
Among them, the geometric mean method (GMM) deserves particular attention.
According to GMM the priority of i-th alternative is formed as the
appropriately rescaled geometric mean of i-th row of the matrix $C$.
Due to its relative simplicity and theoretical properties in recent
times it has gained many supporters.
\begin{example}
Consider a pairwise comparison matrix

\[
C=\left(\begin{array}{cccc}
1 & 1 & 2 & 0.5\\
1 & 1 & 0.25 & 8\\
0.5 & 4 & 1 & 1\\
2 & 0.125 & 1 & 1
\end{array}\right).
\]

Its principal eigenvalue equals $\lambda_{max}\approx5.8875$ and
its principal eigenvector is given by

\[
w_{max}=\left[1.32571,2.0096,1.9849,1\right]^{T}.
\]

The sum of its coordinates equals $6.32021$, so, after normalization,
we obtain a priority vector $\left[0.20976,0.31796,0.31406,0.15822\right]^{T}$.
This determines the order of alternatives: $a_{2},a_{3},a_{1},a_{4}$.
Notice that according to EVM, alternative $a_{2}$ is slightly better
than $a_{3}$. However, since geometric means of the second and third
rows of $C$ are equal, GMM assigns the same weights to both alternatives.
\end{example}

\subsection{Incompleteness}

The priority deriving methods mentioned in the previous section assume
that the set of paired comparisons is complete, i.e., every entry
$c_{ij}$ of $C$ is known and available. In practice, this condition
is not always met. It can happen for many reasons. After taking reciprocity
into account, the number of all possible comparisons for $n$ alternatives
is $n(n-1)/2$. Thus, when the number of alternatives is large comparing
all of them in pairs requires considerable effort. It can not always
be possible, for example, because of limited and expensive work time
of experts. \emph{Harker} \cite{Harker1987amoq} also points out that
an expert, when faced with a comparison between two alternatives $a_{i}$
and $a_{j}$, sometimes would rather not compare them directly. This
may happen when, e.g., they do not yet have a good understanding of
his or her preferences for this particular pair of alternatives. Sometimes
experts evade from the answers, especially when taking a position
on the given comparisons is morally or ethically tricky, e.g., comparing
mortality risk vs. cost. Finally, some data may be lost or damaged. 

In response to the above problems, the methods of calculating the
ranking based on an incomplete set of pairwise comparisons arose.
Probably one of the most popular (and the first one) is the \emph{Harker}
method \cite{Harker1987amoq}. According to the method based on matrix
$C$, a new auxiliary matrix $B=(b_{ij})$ is created where 
\[
b_{ij}=\begin{cases}
c_{ij}, & \text{if \,\,}c_{ij}\,\,\text{exists and}\,\,i\neq j,\\
0, & \text{if \,\,}c_{ij}\,\,\text{does not exist and}\,\,i\neq j,\\
b_{ii}, & \text{if\,\,}i=j,
\end{cases}
\]
and $b_{ii}$ means the number of the unanswered questions in the
i-th row of $C$. Harker has shown that a non-negative quasi-reciprocal
matrix $(B+\textit{Id})$ can be used for calculation priority ranking
as a replacement for an original PC matrix. The natural limitation
of the \emph{Harker} method is that in $C$ there must be a series
of comparisons between every two alternatives $a_{i}$ and $a_{j}$
such that $c_{ik_{1}},c_{k_{1}k_{2}},\ldots,c_{k_{q}j}$ exist. In
other words, every two alternatives must be comparable at least indirectly.
Every matrix $C$ for which the above condition holds is \emph{irreducible}
and every graph $G=(V,E)$ in which the set of vertices $V=\{v_{1},\ldots,v_{n}\}$
correspond to the set of alternatives $a_{1},\ldots,a_{n}$, and the
set of edges $E$ so that there exists the edge $(v_{i},v_{j})$ in
$E$ if $c_{ij}$ is known and defined, is strongly connected \cite{Quarteroni2000nm}.
Let us consider the following example.
\begin{example}
Let $C$ be incomplete PC matrix
\[
C=\left(\begin{array}{ccc}
1 & 3 & ?\\
1/3 & 1 & 3\\
? & 1/3 & 1
\end{array}\right),
\]
hence the \emph{Harker's} auxiliary matrix is
\[
B+\textit{Id}=\left(\begin{array}{ccc}
1 & 3 & 0\\
1/3 & 0 & 3\\
0 & 1/3 & 1
\end{array}\right)+\left(\begin{array}{ccc}
1 & 0 & 0\\
0 & 1 & 0\\
0 & 0 & 1
\end{array}\right),
\]
and
\[
B+\textit{Id}=\left(\begin{array}{ccc}
2 & 3 & 0\\
1/3 & 1 & 3\\
0 & 1/3 & 2
\end{array}\right).
\]
Thus the rescaled ranking vector obtained by EVM is 
\[
w=[0.692,0.23,0.0769]^{T}
\]
which means that the priority of the first alternative is $w(a_{1})=0.692$,
and the second and third: $w(a_{2})=0.23$, $w(a_{3})=0.0769$, correspondingly. 

The corresponding graph has been shown in Fig. \ref{fig:graphC}.

\begin{figure}[tbh]
\begin{centering}
\includegraphics[scale=0.5]{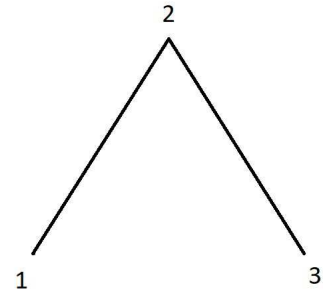}
\par\end{centering}
\caption{The graph of $C$ }

\label{fig:graphC}
\end{figure}
\end{example}

\subsection{Inconsistency}

As $c_{ik}$ represents the results of the comparisons between i-th
and k-th alternative, and $c_{kj}$ expresses the outcome of similitude
between k-th and j-th alternative its natural to expect that $c_{ij}=c_{ik}c_{kj}$.
However,  the entries of the PC matrix $C$ represent the subjective
opinions of experts and due to human imperfection may happen that
$c_{ij}\neq c_{ik}c_{kj}$. Whenever it happens, we will call such
a situation as \emph{inconsistency}. If the difference between $c_{ij}$
and $c_{ik}c_{kj}$ is small or happens rarely, it probably will not
have much impact on the final result. However, if the difference is
large and it happens relatively often, then the results of the pairwise
comparisons may be considered unreliable, and hence, the ranking results
may not be trustworthy. This observation leads to a question about
the degree of inconsistency of PC matrix $C$. A popular way of determining
the level of inconsistency in a set of pairwise comparisons is the
use of inconsistency indexes. Probably the best-known index is one
proposed by \emph{Saaty} in 1977 \cite{Saaty1977asmf}. It is defined
as: 

\[
\textit{CI}=\frac{\lambda_{\textit{max}}-n}{n-1},
\]
where $\lambda_{\textit{max}}$ is the principal eigenvector of $C$,
and $n$ is the number of alternatives. It has been proven that $\textit{CI}$
reaches $0$ when the PC matrix $C$ is fully consistent, and it gets
the higher values, the more inconsistent $C$ is \cite{Saaty1977asmf}.
Since then, many different inconsistency indexes have been created.
A comprehensive overview of the inconsistency indexes can be found
in \cite{Brunelli2013apoi,Brunelli2013iifp,Kulakowski2014tntb}. 

\subsection{Sensitivity}

Another factor that affects the credibility of the ranking is the
sensitivity of the result. By sensitivity we mean here the extent
to which the disturbance of input data can change the final result.
If a little disorder can significantly modify the result, then the
ranking result is unstable and, therefore, not credible (we can not
be sure if the final result is not accidental). Reversely, if reasonably
small changes in the input data do not cause noticeable modifications
of the result, then we can trust that the result obtained is a consequence
of decision-making data deliberately introduced into the system. Simply
put, it can be assumed that the sensitivity can be used to determine
the data quality. The problem is, however, that the sensitivity is
hardly measurable. What does it mean \textquotedbl a small change
in input\textquotedbl ? What change, how to quantify it? What does
it mean ``noticeable modifications of the result''? A person who
wants to deal with the sensitivity analysis must answer all these
questions. For the purpose of this article, we have assumed that the
inconsistency index determines the input data disturbance. To measure
the extent to which the results have been modified, we use two methods:
the \emph{Manhattan} distance \footnote{In \cite{Harker1987ipci} \emph{Harker} used \emph{Chebyshev} distance
$\left\Vert \cdot\right\Vert _{\infty}$for this purpose.} and the rescaled \emph{Kendall tau} distance. 

The \emph{Manhattan} distance between two priority vectors $w$ and
$u$ is defined as follows: 
\begin{equation}
M_{\textit{d}}(w,u)=\sum_{i=1}^{n}\left|w(a_{i})-u(a_{i})\right|.\label{eq:manhattan-distance-eq}
\end{equation}
This metric provides us with information what is the average difference
between two different priorities assigned to the same alternatives.
As all the entries of priority vectors sum up to $1$, the result
$M_{\textit{d}}(w,u)\leq2$. 

Very often the ranking results are interpreted only qualitatively.
This means that the decision makers are interested in who is the winner,
who is in the second and in the third place but not what are the numerical
priorities of alternatives.  Let $O:\mathbb{R}_{+}^{n}\rightarrow\{1,\ldots,n\}^{n}$
be the mapping assigning to every ranking vector $w$ its ordinal
counterpart in such a way that i-th element of $O(w)$ indicates the
position of i-th alternative in the ranking (\ref{eq:ranking_vector}).
For example, if 
\[
w=[0.3,0.5,0.2]^{T}
\]
then its ordinal vector is
\begin{equation}
O(w)=[2,1,3]^{T}.\label{eq:ordinal-vecotr-example}
\end{equation}
Qualitative interpretation of ranking vectors leads to a question
to what extent both: $O(w)$ and $O(u)$ differ from each other. The
answer can be the \emph{Kendall tau} rank distance that counts the
number of pairwise disagreements between two ranking lists \cite{Kendall1938anmo,Fagin2006cpr}.
Let us define \emph{Kendall tau }distance formally:
\begin{multline*}
K_{\textit{d}}(p,q)=\#\left\{ (i,j)\,|\,\,i<j\,\,\text{and}\right.\\
\,\,\left.\textit{sign}(p(a_{i})-p(a_{j}))\neq\textit{sign}(q(a_{i})-q(a_{j}))\right\} 
\end{multline*}
where $p,q$ are ordinal vectors. Since the maximal value of $K_{\textit{d}}(p,q)$
for two n-element vectors is $n(n-1)/2$ it is convenient to use the
rescaled \emph{Kendall} tau distance, i.e. 
\[
K_{\textit{rd}}(p,q)=\frac{2K_{d}(p,q)}{n(n-1)},
\]
so that $0\leq K_{\textit{rd}}(p,q)\leq1$. The rescaled \emph{Kendall
tau} distance is the second method used in the article for the purpose
of measuring discordance between ranking results. Since, vectors produced
by EVM, GMM or \emph{Harker} method are not ordinal before applying
\emph{$K_{\textit{rd}}$} they have to be transformed to their ordinal
counterparts using $O$ mapping. 

Sometimes the \emph{Kendall tau distance} is called a Bubble sort
distance. The reason is that when there are no ties their value represents
the number of swaps that are done by the bubble sort algorithm \cite{Cormen2009ita}
when transforming the first list into the second one.
\begin{example}
Let us consider two ordinal vectors $p=[1,2,4,3]^{T}$ and $q=[3,4,1,2]^{T}$.
It is easy to observe that $K_{\textit{d}}(p,q)=5$ as the discordant
pairs of indices are: $\,(1,3),\,(1,4),\,(2,3),(2,4),(3,4)$. Indeed
there are five binary swaps needed to transform $p$ into $q$. They
are: 
\begin{enumerate}
\item $p=[1,2,4,3]^{T}\rightarrow[1,2,3,4]^{T}$,
\item $[1,2,3,4]^{T}\rightarrow[1,3,2,4]^{T}$,
\item $[1,3,2,4]^{T}\rightarrow[3,1,2,4]^{T}$,
\item $[3,1,2,4]^{T}\rightarrow[3,1,4,2]^{T}$,
\item $[3,1,4,2]^{T}\rightarrow[3,4,1,2]^{T}=q$.
\end{enumerate}
Assuming $n=4$, the rescaled value is $K_{\textit{rd}}(p,q)=\nicefrac{5}{6}$. 
\end{example}

\section{Indices of incompleteness\label{sec:Indices-of-incompleteness}}

\subsection{Incompleteness and sensitivity }

According to EVM, the priority vector meets the equation (\ref{eq:eigenvector-equation}).
In other words, the weight of every alternative $w(a_{i})$ meets
the equation 
\begin{equation}
w(a_{i})=\frac{1}{\lambda_{\textit{max}}}\sum_{j=1}^{n}c_{ij}w(a_{j}).\label{eq:eigenvector-eq-by-element}
\end{equation}
Hence, the priority of one alternative is expressed by the weighted
average of all others alternatives. With this regularity, we also
deal with the case of GMM \cite{Kulakowski2016srot}. The equation
(\ref{eq:eigenvector-eq-by-element}) suggests that the disturbance
of one single element $c_{ij}$, assuming that the other elements
have not changed, should not affect significantly the value of $w(a_{i})$.
However, in the case of an incomplete PC matrix, the relationships
between alternatives are weakened. The priorities of individual alternatives
are determined by fewer expressions in the form $c_{ij}w(a_{j})$
than normally. It suggests that the susceptibility for disturbances
of the rankings calculated based on the incomplete PC matrices is
higher than normal. This, of course, should translate to the usually
higher sensitivity of such decision models. It means that the completeness
of the matrix correlates with the sensitivity of the method. The more
comparisons are available, the less vulnerable the model is. One may
ask whether the number of missing elements is not enough as an index?
To answer this question let us consider the following two PC matrices
with three (six, when the reciprocal elements are taken into account)
missing comparisons. 

\begin{equation}
C_{1}=\left(\begin{array}{ccccc}
1 & c_{12} & ? & ? & ?\\
c_{21} & 1 & c_{23} & c_{24} & c_{25}\\
? & c_{32} & 1 & c_{34} & c_{35}\\
? & c_{42} & c_{43} & 1 & c_{45}\\
? & c_{52} & c_{53} & c_{54} & 1
\end{array}\right),\label{eq:1PCM}
\end{equation}

\begin{equation}
C_{2}=\left(\begin{array}{ccccc}
1 & c_{12} & ? & ? & c_{15}\\
c_{21} & 1 & c_{23} & ? & c_{25}\\
? & c_{32} & 1 & c_{34} & c_{35}\\
? & ? & c_{43} & 1 & c_{45}\\
c_{51} & c_{52} & c_{53} & c_{54} & 1
\end{array}\right).\label{eq:2PCM}
\end{equation}
In the first matrix $a_{1}$ is compared only with $a_{2}$. Thus,
disturbance on $c_{12}$ completely changes the value $w(a_{1})$.
In the second matrix $a_{1}$ is compared with $a_{2}$ and $a_{5}$.
Therefore, the same disturbance on $c_{12}$ will have less impact
on the priority $w(a_{1})$.  In Section \ref{sec:Montecarlo-section}
this intuition will be confirmed by the Montecarlo experiment. The
above consideration leads us to the conclusion that the completeness
index, that would be useful in determining the sensitivity of the
decision model, should also take into account the arrangement of missing
comparisons. 

\subsection{$\alpha$-index}

In $n\times n$ PC matrix a single alternative can be compared with
at most $n-1$ other alternatives. Therefore, the maximal value of
$\textit{outdeg}(a_{i})$ for $i=1,\ldots,n$ is $n-1$ (see Def.
\ref{def:Let-the-output}). Similarly, the number of missing comparisons
is given by $n-1-\textit{outdeg}(a_{i})$. Because the desired behavior
is that the newly constructed index should be higher for $C_{1}$
than for $C_{2}$ the higher value of the expression $n-1-\textit{outdeg}(a_{i})$
for some particular $i$ should contribute more to the value of the
index than two or more smaller expressions. To achieve this let us
raise the expression $\left(n-1-\textit{outdeg}(a_{i})\right)^{\alpha}$
to a positive real number $\alpha>1$. Thus, the expression 
\[
S_{\alpha}(C)=\sum_{i=1}^{n}\left(n-1-\textit{outdeg}(a_{i})\right)^{\alpha}
\]
combines two features together. It raises when the number of missing
comparisons increases and providing that there are two matrices of
the same size and with the same number of missing comparisons it is
higher for this matrix that has larger irregularities in the distribution
of missing values. Let us compute the mean of missing values raised
to $\alpha>1$. As a result, we get the formula: 

\begin{equation}
\frac{1}{n}S_{\alpha}(C)\label{eq:eigenvector-eq-by-element-3}
\end{equation}
which preserves both important features and its value is bounded and
varies within the range $[0,(n-1)^{\alpha}]$. Hence, in order to
get the final form of the index let us divide (\ref{eq:eigenvector-eq-by-element-3})
by $(n-1)^{\alpha}$, i.e. 

\[
\textit{IId}_{\alpha}(C)=\frac{\frac{1}{n}S_{\alpha}(C)}{(n-1)^{\alpha}}.
\]
It is clear that $0\leq\textit{IId}_{\alpha}\leq1$. When the PC matrix
is fully incomplete, i.e. there are no comparisons between alternatives,
$\textit{IId}_{\alpha}(C)$ is $0$. Reversely, if $C$ is complete,
i.e. all the alternatives are defined, $\textit{IId}_{\alpha}(C)$
equals $1$. Providing that the PC matrix is reciprocal, every alternative
has to be compared with at least one different alternative. The maximal
value of $\textit{IId}_{\alpha}(C)$ that allows to create the ranking
is reached when just one alternative is compared with all the others.
Then it is given by $\frac{n-1}{n}\cdot\left(\frac{n-2}{n-1}\right)^{\alpha}$.
Condition 
\[
\textit{IId}_{\alpha}(C)\leq\frac{n-1}{n}\cdot\left(\frac{n-2}{n-1}\right)^{\alpha}
\]
 is necessary but it is not sufficient. Hence, there may exist PC
matrices for which $\textit{IId}_{\alpha}$ is smaller than $\frac{n-1}{n}\cdot\left(\frac{n-2}{n-1}\right)^{\alpha}$
but, in spite of this, one can not create the ranking. 
\begin{example}
Consider matrices $C_{1}$ and $C_{2}$ given by (\ref{eq:1PCM})
and (\ref{eq:2PCM}). Let us calculate their $2$-indices:

\begin{multline*}
\textit{IId}_{2}(C_{1})=\frac{\frac{1}{5}\sum_{i=1}^{5}\left(4-\textit{outdeg}(a_{i})\right)^{2}}{16}=\frac{9+1+1+1}{80}=\\
=0.15.
\end{multline*}

\begin{multline*}
\textit{IId}_{2}(C_{2})=\frac{\frac{1}{5}\sum_{i=1}^{5}\left(4-\textit{outdeg}(a_{i})\right)^{2}}{16}=\frac{4+1+1+4}{80}=\\
=0.125.
\end{multline*}
\end{example}
As we can see the index of the first matrix is greater than the index
of the second one, which reflects the fact that the distribution of
the missing items in the rows of $C_{2}$ is more aligned than in
$C_{1}$. However, both indices are quite small, as both matrices
lack of only $6$ elements (out of $20$).

\subsection{$\beta$ index }

According the old adage ``a chain is only as strong as its weakest
link''. Following this common sense observation the second index
does not consider the average number of missing comparisons for all
alternatives but it focuses on the maximum number of missing comparisons
for a single alternative: 

\[
M(C,\beta)=\left(\max_{i=1,\ldots,n}\left(n-1-\textit{outdeg}(a_{i})\right)\right)^{\beta}.
\]
Because we can not omit the total number of comparisons the ``weakest
link'' in the form of the above formula has to be combined with the
sum:
\[
S(C)=\sum_{i=1}^{n}\left(n-1-\textit{outdeg}(a_{i})\right).
\]
Thus, the proposed index gets the form:

\[
\textit{II}_{\beta}(C)=\frac{M(C,\beta)S(C)}{n(n-1)^{1+\beta}},
\]
where the multiplier $1/n(n-1)^{1+\beta}$ is introduced only for
the purpose of fitting the index value to the segment $[0,1]$. It
is easy to observe that $\textit{II}_{\beta}(C)$ is $0$ when the
matrix $C$ is complete. Reversely, $\textit{II}_{\beta}(C)=1$ if
there are no defined values in the matrix except its diagonal. 
\begin{example}
Similarly as before let us consider $C_{1}$ and $C_{2}$ given by
(\ref{eq:1PCM}) and (\ref{eq:2PCM}). Their $\beta$ indices (where
$\beta=1$) are

\begin{multline*}
\textit{II}_{\beta}(C_{1})=\frac{\max\left\{ 3,1,1,1,0\right\} \cdot\left(3+1+1+1+0\right)}{80}=\\
=\frac{9}{40}=0.225.
\end{multline*}

\begin{multline*}
\textit{II}_{\textit{\ensuremath{\beta}}}(C_{2})=\frac{\max\left\{ 2,1,1,2,0\right\} \cdot\left(2+1+1+2+0\right)}{80}=\\
=\frac{3}{20}=0.15.
\end{multline*}
\end{example}
Similarly, as in the case of $\alpha$-index, the matrix $C_{1}$
gets the higher values of the index than the matrix $C_{2}$. Both
values, however, are quite small as only six elements (out of $20$)
are missing. 

\subsection{Tree index}

As \cite{Koczkodaj2015pcs} shows, the existence of a spanning tree
in the graph associated with an incomplete pairwise comparison matrix
is a necessary condition to generate its missing values, and what
follows, to create the ranking. Of course, the more spanning trees
we have, the more reliable the data we obtain. The Cayley's formula
\cite{Cayley1889atot} states that the number of all spanning trees
in a complete graph with $n$ vertices is equal to $n^{n-2}$. If
we consider a complete PC matrix, its incompleteness index should
be equal to 0. The index should raise with the reduction of the number
of spanning trees. However, as we remove the matrix entries one by
one, the number of trees decreases exponentially from $n^{n-2}$ to
$0$. To slow down its drop occurring when we remove the PC matrix
elements, it is desirable to divide it over $n^{n-2}$ and apply the
$n-2$ root to the ratio. These simple observations lead us to the
definition of an alternative incompleteness indicator, which we will
call the tree index.
\begin{defn}
The tree-index of a pairwise comparison matrix $C$ is defined by
the formula

\[
TI(C)=1-\frac{NT(C)^{\frac{1}{n-2}}}{n},
\]
where $NT(C)$ denotes the number of the spanning trees in a graph
associated with the matrix $C$.
\end{defn}
\begin{rem}
Notice that $TI(C)=0$ if and only if $G_{C}$ is complete i.e. $C$
has got all elements. On the other hand, $TI(C)=1$ if and only if
$G_{C}$ is disconnected, which means that we cannot create a priority
vector based on the elements of $C$.
\end{rem}
According to the Kirchoff's Theorem \cite{Maurer1976mgos} the number
of spanning trees in a connected graph $G$ with $n$ vertices $v_{1},\ldots,v_{n}$
can be computed as any cofactor of the Laplacian matrix $L(G)=[l_{ij}]$
of $G$, whose elements are given by the formula:

\[
l_{ij}=\begin{cases}
deg(v_{i}), & \mbox{if }i=j,\\
-1, & \mbox{if \ensuremath{i\neq j} and \ensuremath{v_{i}} is connected with \ensuremath{v_{j},}}\\
0, & \mbox{otherwise}.
\end{cases}
\]

\begin{example}
Once more, consider matrices $C_{1}$and $C_{2}$ given by (\ref{eq:1PCM})
and (\ref{eq:2PCM}). The corresponding graphs $G_{C_{1}}$ and $G_{C_{2}}$
are given in Fig. \ref{Fig:G12}. 

\begin{figure}[tbh]
\begin{centering}
\includegraphics[scale=0.5]{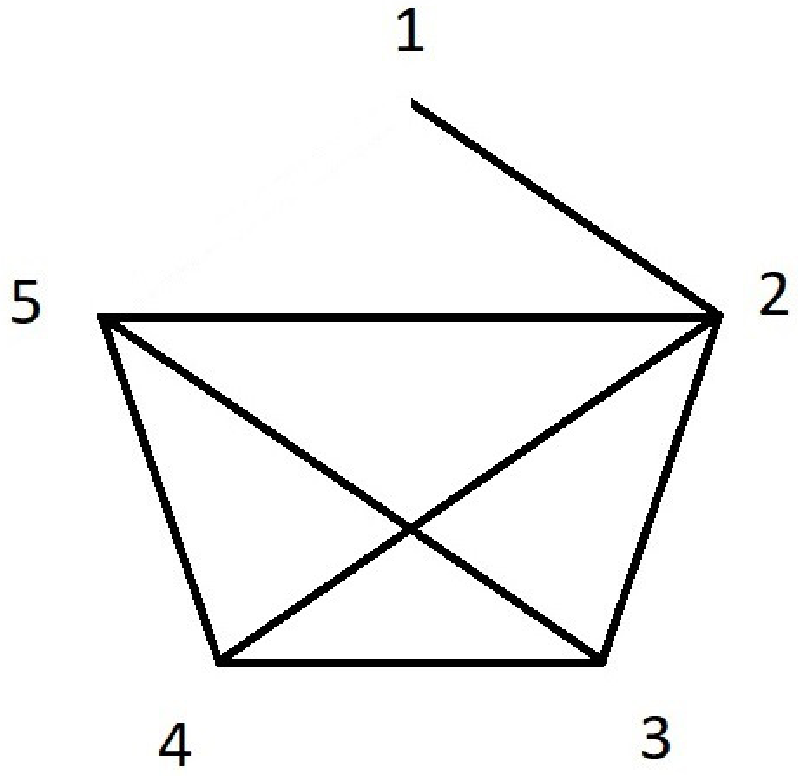}\includegraphics[scale=0.5]{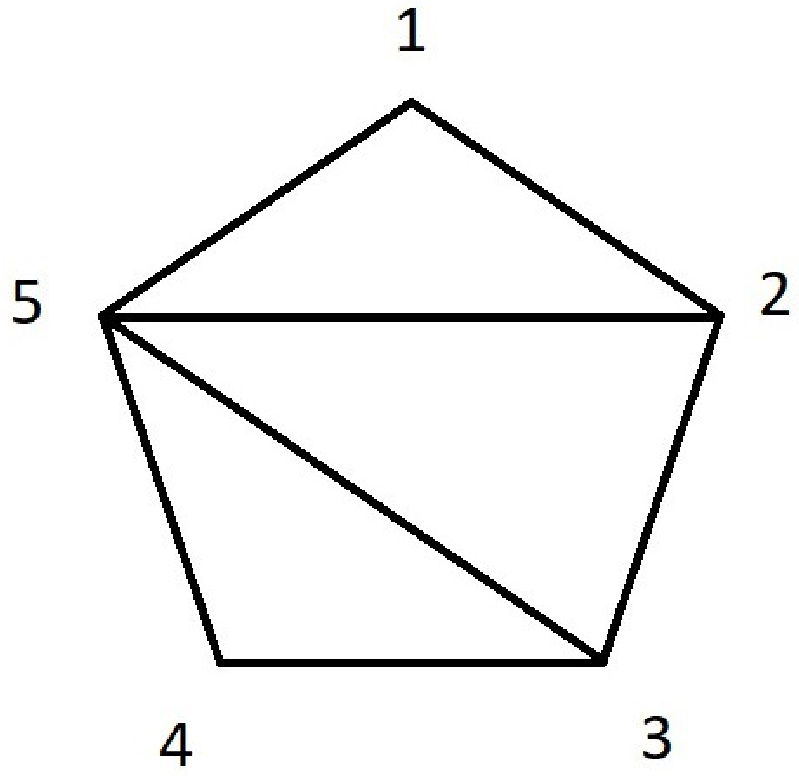}
\par\end{centering}
\caption{Graphs $G_{C_{1}}$and $G_{C_{1}}$}
\label{Fig:G12}
\end{figure}

Their Laplacian matrices are as follows:

\[
L(G_{C_{1}})=\left(\begin{array}{ccccc}
1 & -1 & 0 & 0 & 0\\
-1 & 4 & -1 & -1 & -1\\
0 & -1 & 3 & -1 & -1\\
0 & -1 & -1 & 3 & -1\\
0 & -1 & -1 & -1 & 3
\end{array}\right)
\]

and

\[
L(G_{C_{2}})=\left(\begin{array}{ccccc}
2 & -1 & 0 & 0 & -1\\
-1 & 3 & -1 & 0 & -1\\
0 & -1 & 3 & -1 & -1\\
0 & -1 & -1 & 2 & -1\\
-1 & -1 & -1 & -1 & 4
\end{array}\right).
\]

Let us compute the cofactors of the left upper elements of the above
matrices:

\[
G_{C_{1}11}=(-1)^{2}\cdot\left|\begin{array}{cccc}
4 & -1 & -1 & -1\\
-1 & 3 & -1 & -1\\
-1 & -1 & 3 & -1\\
-1 & -1 & -1 & 3
\end{array}\right|=16,
\]

\[
G_{C_{1}11}=2\cdot(-1)^{2}\cdot\left|\begin{array}{cccc}
3 & -1 & 0 & -1\\
-1 & 3 & -1 & -1\\
0 & -1 & 2 & -1\\
-1 & -1 & -1 & 4
\end{array}\right|=42.
\]

According to the Kirchoff's Theorem, graphs $G_{C_{1}}$ and $G_{C_{2}}$
include $16$ and, respectively, $42$ spanning trees. Since $n=5$,
the tree incompleteness indices of $C_{1}$and $C_{2}$ are 

\[
T(C_{1})=1-\frac{\sqrt[3]{16}}{5}=0.496,
\]

\[
T(C_{2})=1-\frac{\sqrt[3]{42}}{5}=0.305.
\]

Again, the index of $C_{1}$ is higher than the index of $C_{2}$,
which reflects the fact that removing six elements of a $5\times5$
matrix reduces the number of the respective graph's spanning trees,
which lowers the reliability of the resulting priority vector.
\end{example}
It is important to point out that both indices may be useful as measures
of incompleteness. The first one measures the location of a matrix
on the line between full and (almost) empty (i.e., having only 1s
on the main diagonal) matrices. The latter reflects how far a matrix
is from matrices which are useless for ordering the alternatives. 

\subsection{Compound indices}

As all the indices have the same domain (PC matrices) and codomain
$[0,1]\subset\mathbb{R}_{+}$, then their product will also be a function
with the same domain and codomain. It allows us to combine one index
with the other to obtain the desirable properties of both. In this
context, an interesting proposal seems to be combining $\alpha$ and
\textbf{$\beta$ }indices. Thus, let us define a compound $\alpha,\beta$-index
as follows: 
\[
\textit{II}_{\alpha,\beta}(C)=\textit{II}_{\alpha}(C)\cdot\textit{II}_{\beta}(C)
\]
As it will turn out in the Section \ref{sec:Montecarlo-section},
this product allows us to combine together a dynamics of average sensitivity
represented by $\textit{II}_{\alpha}$ together with differences in
sensitivity resulting from different arrangements of missing pairwise
comparisons. This second feature seems to be better represented by
$\textit{II}_{\beta}$. 

\section{Properties of incompleteness indices - a numerical study\label{sec:Montecarlo-section} }

\subsection{Relationship between incompleteness, inconsistency and sensitivity\label{subsec:Relationship-between-incompleten}}

An entirely consistent matrix is resistant to reducing the set of
paired comparisons. That is because it suffices to compare one alternative
with another already ranked to precisely determine the ranking of
the former. Hence, as long as it is possible to compute the ranking,
i.e., the PC matrix is \emph{irreducible,} the calculated ranking
is the same regardless of which comparisons are missing. However,
if a PC matrix is inconsistent, missing comparisons start to matter. 

In order to investigate the impact of inconsistency and incompleteness
to the sensitivity we randomly prepare $1000$ complete and consistent
PC matrices $\mathcal{C}=C_{1},\ldots,C_{1000}$. Then every matrix
from $\mathcal{C}$ was disturbed so that we obtain $41$ sets $\mathcal{C}^{1},\ldots,\mathcal{C}^{41}$
of matrices with the increasing average inconsistency $\textit{CI\ensuremath{_{\textit{avg}}}}$
given as 
\[
\textit{CI\ensuremath{_{\textit{avg}}(\mathcal{C}^{j})=}\ensuremath{\frac{1}{n}\sum_{i=1}^{n}}\textit{CI}}(C_{i}).
\]
The average of inconsistencies of those groups starts from $\textit{CI}_{\textit{avg}}(\mathcal{C}^{1})=0.001$,
$\textit{CI}_{\textit{avg}}(C^{2})=0.004$, $\textit{CI}_{\textit{avg}}(C^{3})=0.008$
and finally they reach $\textit{CI}_{\textit{avg}}(C^{41})=0.385$.
Next, we extend every $\mathcal{C}^{j}$ by adding irreducible incomplete
matrices randomly obtained from those originally located\footnote{As irreducible $n\times n$ matrix must have at least $n-1$ comparisons
(we are counting only comparisons over the diagonal) then, for every
inconsistent matrix $C\in\mathcal{C}^{j}$, we generate $n(n-1)/2\,-\,(n-1)=(n^{2}-3n+2)/2$
incomplete matrices.} there. Let us denote the extended $\mathcal{C}^{j}$ by $\widehat{\mathcal{C}}^{j}$
and its elements by $C_{i}^{j,k}\in\widehat{\mathcal{C}}^{j}$, where
$k$ means the number of missing comparisons and $i$ indicates the
consistent PC matrix $C_{i}\in\mathcal{C}$ from which $C_{i}^{j,k}$
originated. For every $C_{i}^{j,k}$ we compute incompleteness indices
$\textit{II}_{\alpha}(C_{i}^{j,k}),\textit{II}_{\beta}(C_{i}^{j,k})$
and $\textit{TI}(C_{i}^{j,k})$, the measures of sensitivity i.e.
Kendall distance $K_{\textit{\textit{rd}}}(w(C_{i}),w(C_{i}^{j,k}))$
and the Manhattan distance $M_{\textit{d}}(w(C_{i}),w(C_{i}^{j,k}))$. 

\begin{figure}[tbh]
\begin{centering}
\subfloat[Incompleteness given as $\textit{II}_{\alpha}$ for $\alpha=1.5$]{\begin{centering}
\includegraphics[scale=0.35]{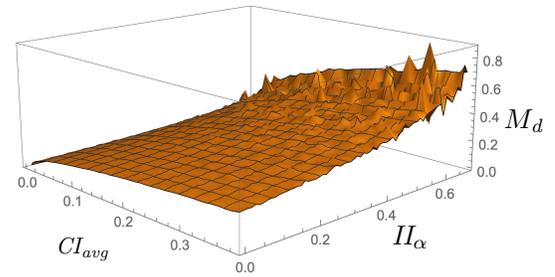}
\par\end{centering}

\label{fig:SensitivityManhattan1-a}}
\par\end{centering}
\begin{centering}
\subfloat[Incompleteness given as $\textit{II}_{\beta}$]{\begin{centering}
\includegraphics[scale=0.3]{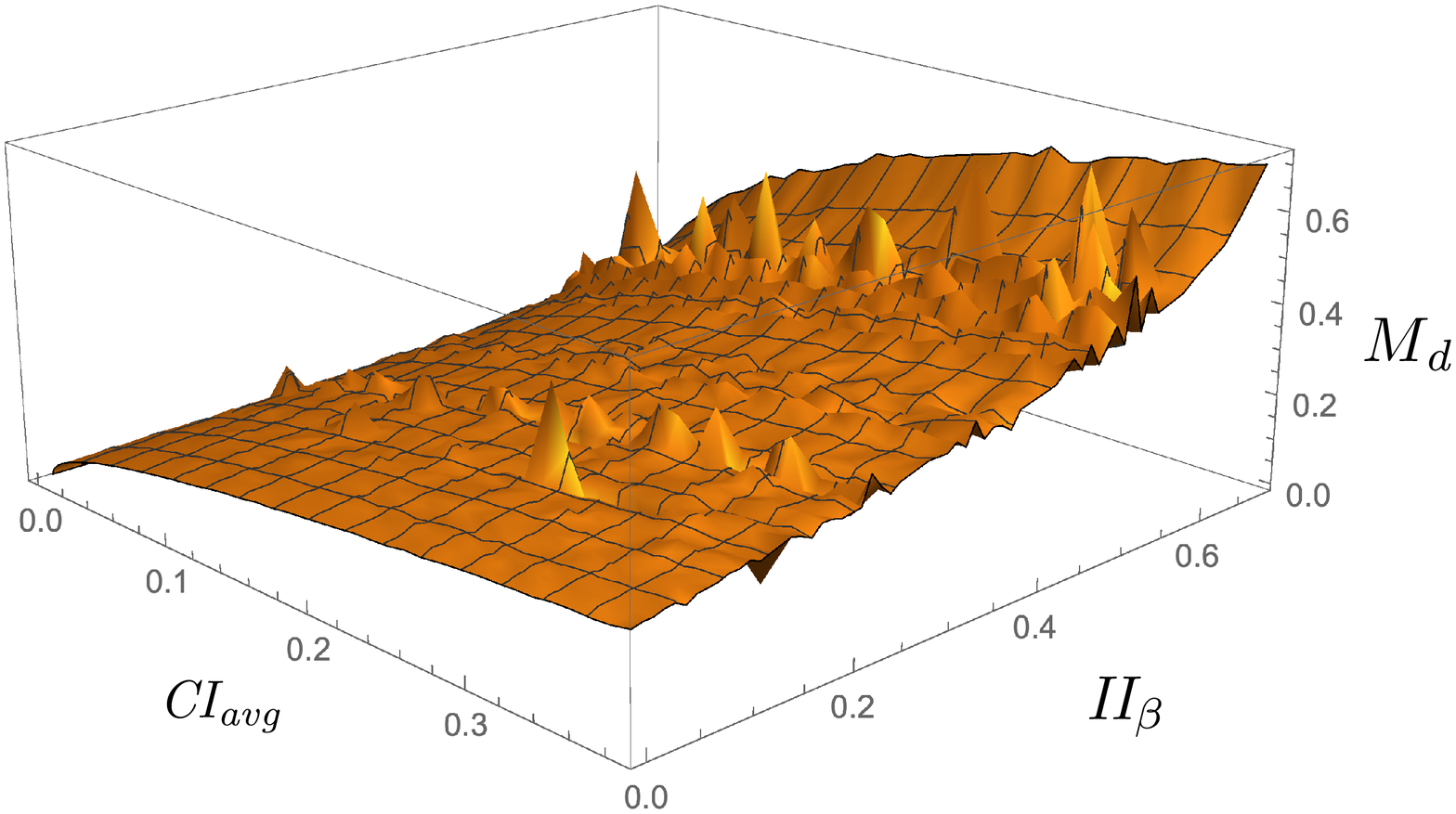}
\par\end{centering}
\label{fig:SensitivityManhattan1-b}}
\par\end{centering}
\begin{centering}
\subfloat[Incompleteness given as $\textit{TI}$]{\begin{centering}
\includegraphics[scale=0.25]{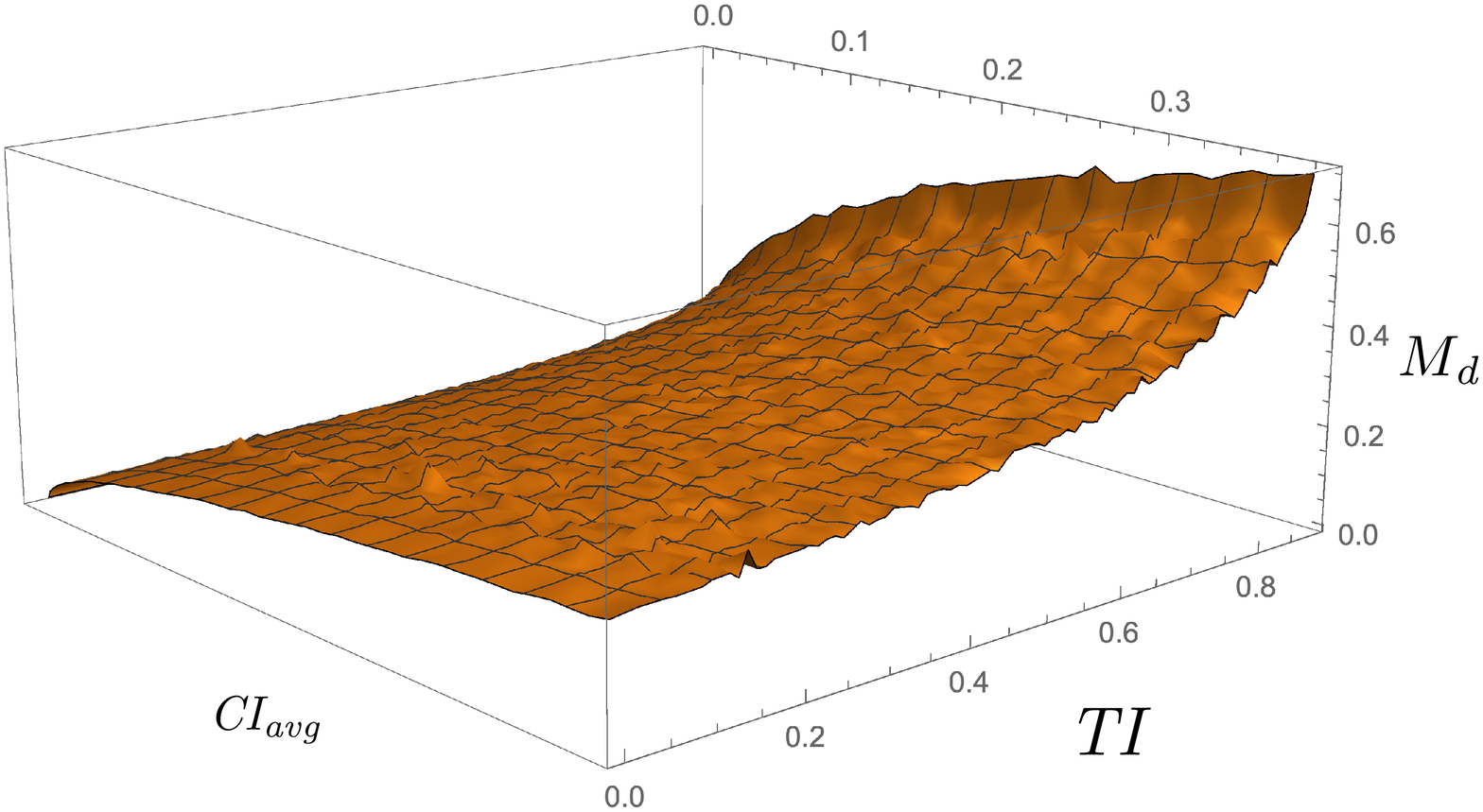}
\par\end{centering}

\label{fig:SensitivityManhattan1-c}}
\par\end{centering}
\caption{Relationship between average consistency level, incompleteness and
sensitivity given as the average Manhattan distance between rankings
obtained from consistent and inconsistent (and incomplete) $9\times9$
matrices. }

\label{fig:SensitivityManhattanFig}
\end{figure}

In the Figure \ref{fig:SensitivityManhattanFig} we can see the relationship
between average value of sensitivity for matrices $C_{i}^{j,k}$ with
the given average inconsistency $CI\ensuremath{_{\textit{avg}}(\mathcal{C}^{j})}$
and the average incompleteness given in the form of the three indices
$\textit{II}_{\alpha}(C_{i}^{j,k})$, $\textit{II}_{\beta}(C_{i}^{j,k})$
and $\textit{TI}(C_{i}^{j,k})$. When the considered PC matrices are
consistent i.e. $CI\ensuremath{_{\textit{avg}}(\mathcal{C}^{j})}=0$
then also the resulting rankings do not depend on incompleteness.
The distance between rankings obtained from consistent complete and
incomplete matrices is $0$. However, when inconsistency starts increasing,
the impact of incompleteness becomes apparent.

The increase of both inconsistency and incompleteness translates to
the increase of the average Manhattan distance. For very small values
of inconsistency ($\textit{CI}_{\textit{avg}}\approx0.001$) the Manhattan
distance is about $0.01$ and following the increase of $\textit{II}_{\alpha}$
it takes values near $0.04$. For the larger values e.g. $\textit{CI}_{\textit{avg}}\approx0.11$
the value of $M_{d}$ ranges between $0.1$ and $0.4$, and similarly
for $\textit{CI}_{\textit{avg}}\approx0.38$ the average values of
$M_{d}$ are between $0.2$ and $0.8$. This observation indicates
that the highly incomplete PC matrices are almost four times more
vulnerable to the random disturbances than the complete matrices.
As the maximal possible value of the Manhattan distance for vectors
whose elements add up to 1 is 2, the value of $M_{d}=0.4$ means that
this index reaches $20\%$ of its maximal value. The similar behavior
can be observed for the other two indices: $\textit{II}_{\beta}$
and $\textit{TI}$ (Figs. \ref{fig:SensitivityManhattan1-b} and \ref{fig:SensitivityManhattan1-c}). 

\begin{figure}[tbh]
\begin{centering}
\subfloat[Incompleteness given as $\textit{II}_{\alpha}$ (for $\alpha=1.5$) ]{\begin{centering}
\includegraphics[scale=0.3]{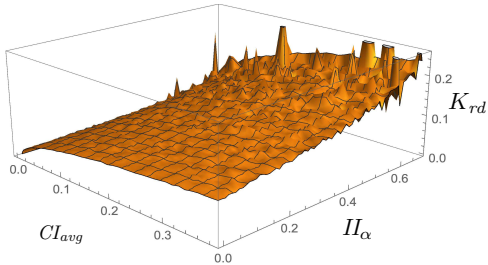}
\par\end{centering}

\centering{}\label{fig:SensitivityKendall-a}}
\par\end{centering}
\begin{centering}
\subfloat[Incompleteness given as $\textit{II}_{\beta}$ ]{\begin{centering}
\includegraphics[scale=0.3]{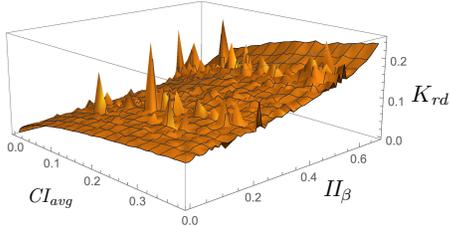}
\par\end{centering}

\centering{}\label{fig:SensitivityKendall-b}}
\par\end{centering}
\begin{centering}
\subfloat[Incompleteness given as $\textit{TI}$]{\begin{centering}
\includegraphics[scale=0.3]{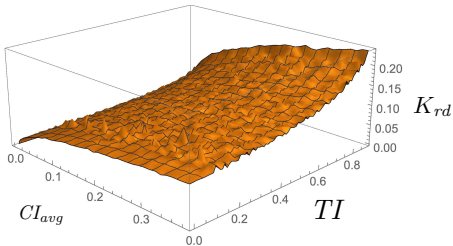}
\par\end{centering}

\centering{}\label{fig:SensitivityKendall-c}}
\par\end{centering}
\caption{Relationship between average consistency level, incompleteness and
sensitivity given as the average Kendall distance between rankings
obtained from consistent and inconsistent (and incomplete) $9\times9$
matrices. }

\label{fig:SensitivityKendallFig}
\end{figure}

The values of \emph{Kendall} distance reveals the similar properties
(Figs. \ref{fig:SensitivityKendall-a}, \ref{fig:SensitivityKendall-b}
and \ref{fig:SensitivityKendall-c}). When the inconsistency is small
($\textit{CI}_{\textit{avg}}\approx0.001$) the average values of
\emph{Kendall} distance are spanned between $0.005$ and $0.025$
for all indices of incompleteness. Then for moderately inconsistent
matrices ($\textit{CI}_{\textit{avg}}\approx0.11$) they range between
$0.05$ and $0.15$, then for ($\textit{CI}_{\textit{avg}}\approx0.38$)
the values of Kendall index go through $0.09$ to $0.25$. 

It means that for PC matrices with the reasonably high inconsistency
we may expect that $25\%$ or more pairs may randomly change their
order. Similarly as before the incompleteness may significantly increase
(from three to four times) the sensitivity of the PC method. 

\subsection{Impact of the distribution of missing comparisons to the sensitivity\label{subsec:Impact-of-the}}

We may suppose that the more missing comparisons to the given alternative
the more vulnerable its weight and the position in the ranking.  In
the extreme case, if the given alternative $a_{i}$ is compared to
only one other alternative $a_{j}$, i.e., except $c_{ij}$, where
$i\neq j$ all other values in the i-th row and j-th column of $C$
are undefined, the ranking of $a_{i}$ depends primarily on $c_{ij}$.
Any disturbance of $c_{ij}$ can translate into significant changes
in the weight of the i-th alternative. On the opposite case, the missing
comparisons are evenly distributed between alternatives. It ensures
the relative safety of each alternative, providing of course, that
the number of missing alternatives is not too high. The above observations
allow us to indicate an example of the regular and the irregular PC
matrix with a fixed number of missing comparisons. 

Let us number the selected entries in the $n\times n$ PC matrix in
such a way that in the first row $c_{13}$ corresponds to $1$, $c_{14}-2$
and $c_{1,n}$ has assigned number $n-2$. Similarly, in the second
row $c_{24}$ gets the number $n-1,$ $c_{25}-n$ and the last element
in the row $c_{2,n}$ gets $2n-4$. Finally, the last element $c_{n-1,n}$
gets the number $(n^{2}-3n+2)/2$. Elements directly above the diagonal
are not indexed (the above numbering scheme has been shown in the
form of a matrix $C_{w}$). 

\[
C_{w}=\left(\begin{array}{cccccc}
1 & c_{12} & c^{(1)} & \cdots & \cdots & c^{(n-2)}\\
 & 1 & c_{23} & c^{(n-1)} & \cdots & c^{(2n-4)}\\
 &  & \ddots & \ddots & \cdots & \vdots\\
 &  &  & \ddots & c_{n-2,n-1} & c^{\left(\frac{n^{2}-3n+2}{2}\right)}\\
 &  &  &  & 1 & c_{n-1,n}\\
 &  &  &  &  & 1
\end{array}\right)
\]
Then, in order to prepare the highly irregular (and highly sensitive)
matrix with $x$ missing comparisons it is enough to remove comparisons
with assigned numbers from $1$ to $x$ and their counterparts below
the diagonal. For example, the highly irregular $7$ by $7$ PC matrix
with $9$ missing comparisons may look like: 
\[
C_{w}^{(9)}=\left(\begin{array}{ccccccc}
1 & c_{12} & ? & ? & ? & ? & ?\\
c_{21} & 1 & c_{23} & ? & ? & ? & ?\\
? & c_{32} & 1 & c_{34} & c_{35} & c_{36} & c_{37}\\
? & ? & c_{43} & 1 & c_{45} & c_{46} & c_{47}\\
? & ? & c_{53} & c_{54} & 1 & c_{56} & c_{57}\\
? & ? & c_{63} & c_{64} & c_{65} & 1 & c_{67}\\
? & ? & c_{73} & c_{74} & c_{75} & c_{76} & 1
\end{array}\right)
\]
For the purpose of creating the matrices with the most even distribution
of missing values we use another numbering scheme. Let assign number
$1$ to $c_{13}$, $2$ to $c_{24}$, $3$ to $c_{35}$, and $n-2$
to $c_{n-2,n}$. The number $n-1$ be assigned to $c_{14}$, $n$
to $c_{25}$ and finally $2n-4$ to $c_{n-3,n}$. The last numbered
element is $c_{1n}$ with value of index $(n^{2}-3n+2)/2$ (the regular
numbering scheme is shown as the matrix $C_{b}$)

\[
C_{b}=\left(\begin{array}{cccccc}
1 & c_{12} & c^{(1)} & c^{(n-1)} & \cdots & c^{(\frac{n^{2}-3n+2}{2})}\\
 & 1 & c_{23} & c^{(2)} & \ddots & \vdots\\
 &  & \ddots & \ddots & \ddots & \vdots\\
 &  &  & \ddots & c_{n-2,n-1} & c^{\left(n-2\right)}\\
 &  &  &  & 1 & c_{n-1,n}\\
 &  &  &  &  & 1
\end{array}\right)
\]
For example, the regular $7$ by $7$ PC matrix with $9$ missing
comparisons is as follows:

\[
C_{b}^{(9)}=\left(\begin{array}{ccccccc}
1 & c_{12} & ? & ? & c_{15} & c_{16} & c_{17}\\
c_{21} & 1 & c_{23} & ? & ? & c_{26} & c_{27}\\
? & c_{32} & 1 & c_{34} & ? & ? & c_{37}\\
? & ? & c_{43} & 1 & c_{45} & ? & ?\\
c_{51} & ? & ? & c_{54} & 1 & c_{56} & ?\\
c_{61} & c_{62} & ? & ? & c_{65} & 1 & c_{67}\\
c_{71} & c_{72} & c_{73} & ? & ? & c_{76} & 1
\end{array}\right)
\]
It is easy to observe that in $C_{b}^{(9)}$ all alternatives have
two missing comparisons (so each of them is compared with the three
others), while in $C_{w}^{(9)}$ alternative $a_{1}$ is compared
only with $a_{2}$ and $a_{2}$ is compared only with $a_{1}$ and
$a_{3}$. In the worst case the disturbances of $c_{12}$ and $c_{23}$
may lead to significant weight changes of $a_{1}$ and $a_{2}$. 

The question arises to what extent the regular and irregular distribution
of missing comparisons translates to the measured sensitivity, and
of course to the values of incompleteness indices. In order to answer
these questions, we prepared $1000$ random inconsistent and incomplete
PC matrices $9\times9$ with the average inconsistency $\textit{CI}\approx0.1$
then we removed their elements according to both: the regular $C_{b}^{(i)}$
and the irregular $C_{w}^{(i)}$ pattern subsequently assuming\footnote{Note that for $n=9$ we get $\frac{n^{2}-3n+2}{2}=28$. }
$i=0,1,2,\ldots,28$ missing elements. Then we measured the average
distance of the ranking vectors obtained from $C_{b}^{(i)}$ and $C_{w}^{(i)}$
and complete and not disturbed matrix, and, similarly, we computed
the average value of all four indices including the compound $\alpha,\beta$-index. 

\begin{figure}[tbh]
\subfloat[Manhattan distance]{\begin{centering}
\includegraphics[scale=0.3]{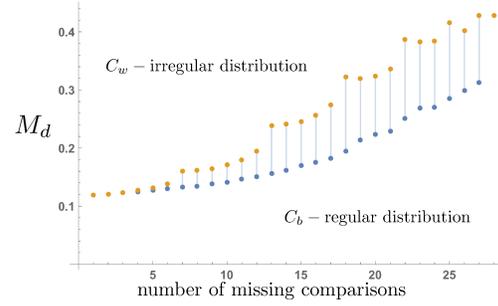}
\par\end{centering}
\label{fig:impact-manhattan}}

\subfloat[Kendall distance]{\begin{centering}
\includegraphics[scale=0.3]{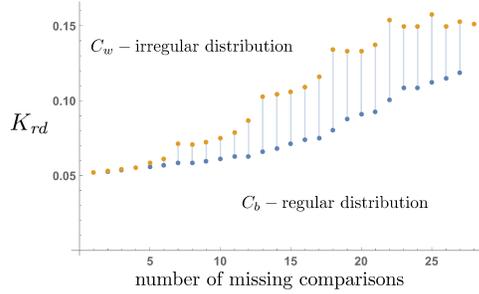}
\par\end{centering}
\label{fig:impact-kendall}}

\caption{Impact of the distribution of missing comparisons (the lower the better),
measured in the group of random PC matrices $9\times9$ with the average
inconsistency $\textit{CI}\approx0.1$.}

\label{fig:impact-indices}
\end{figure}

In the Figures \ref{fig:impact-manhattan} and \ref{fig:impact-kendall}
we can see two plots. The lower plot on both figures represents the
average sensitivity of incomplete PC matrices with the missing values
distributed according to the $C_{b}$ scheme. The upper plot corresponds
to the average sensitivity of incomplete PC matrices with the missing
values distributed according to $C_{w}$. Both plots look quite similar.
They grow as the number of missing comparisons increases, but the
plot corresponding to the irregular incompleteness scheme grows faster.
It is interesting to note that starting from thirteen missing comparisons
the difference in sensitivity between matrices in the form $C_{b}$
and $C_{w}$ reaches almost $40\%$. It shows how important for sensitivity
the distribution of missing comparisons is. 

\begin{figure}[tbh]
\begin{centering}
\subfloat[$\alpha$-index (with $\alpha=1.5$)]{\begin{centering}
\includegraphics[scale=0.3]{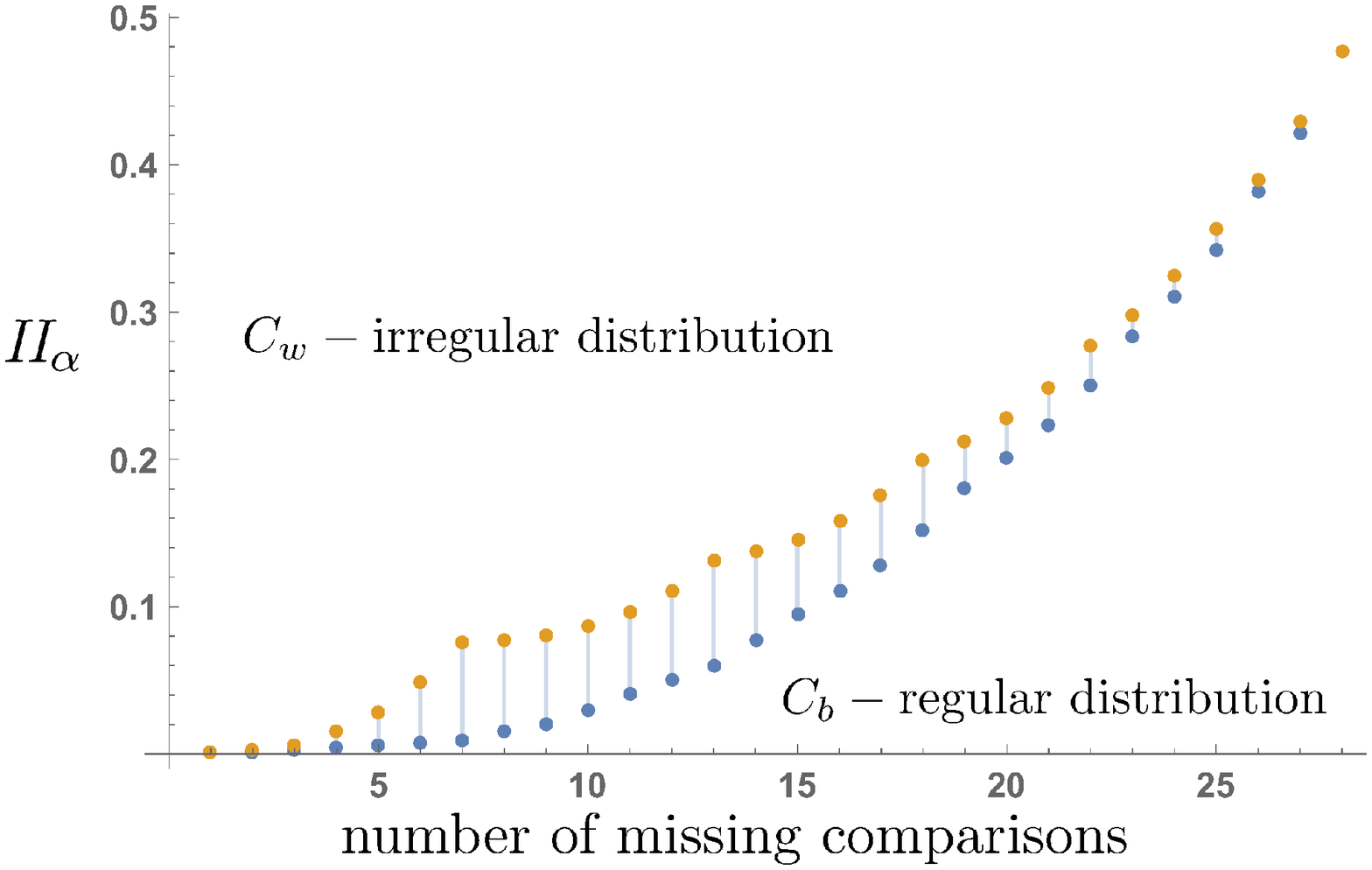}
\par\end{centering}
\label{fig:impact-alpha-index}}
\par\end{centering}
\begin{centering}
\subfloat[$\beta$-index, (with $\beta=1$)]{\begin{centering}
\includegraphics[scale=0.3]{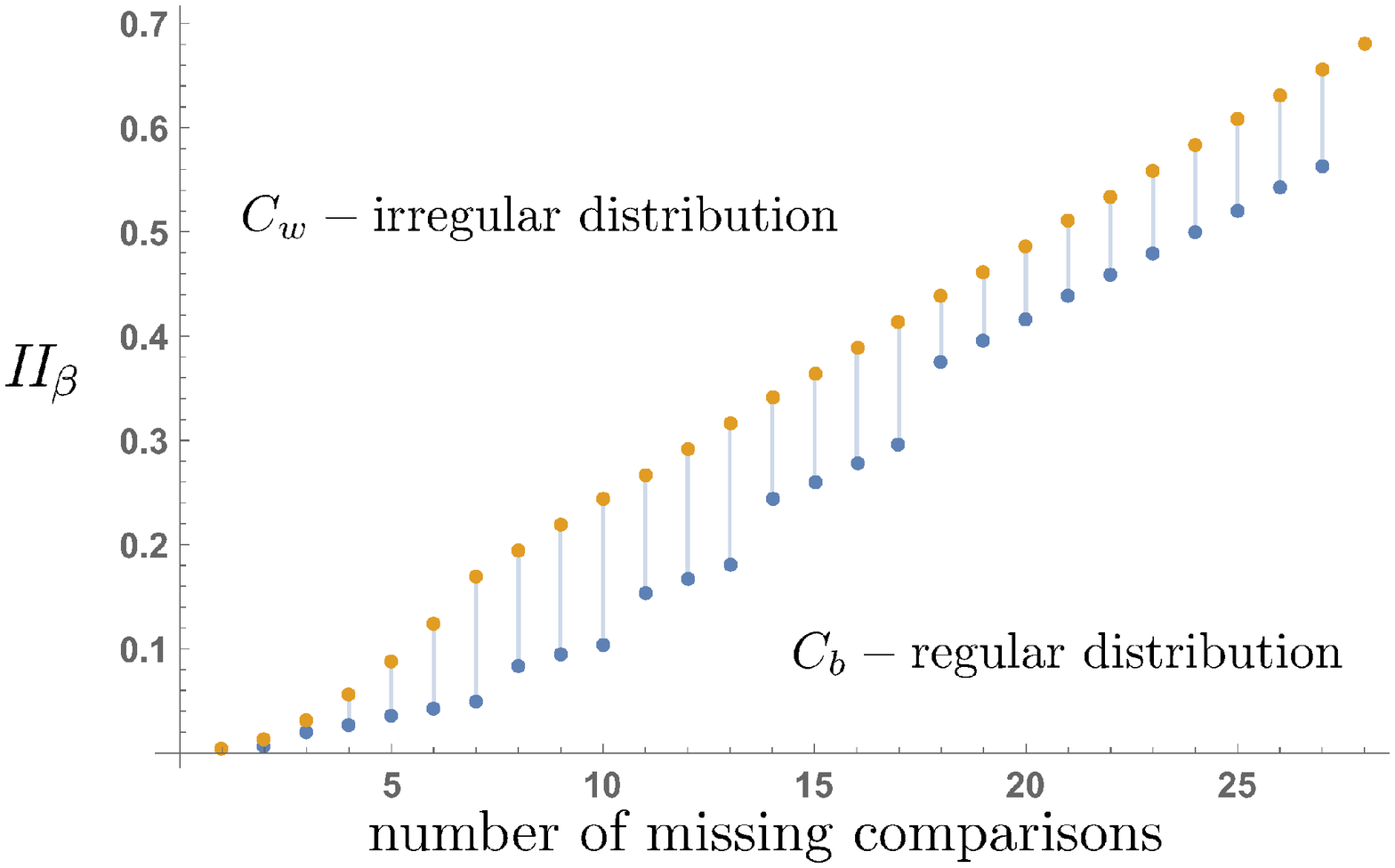}
\par\end{centering}

\label{fig:impact-max-index}}
\par\end{centering}
\begin{centering}
\subfloat[Tree index]{\begin{centering}
\includegraphics[scale=0.3]{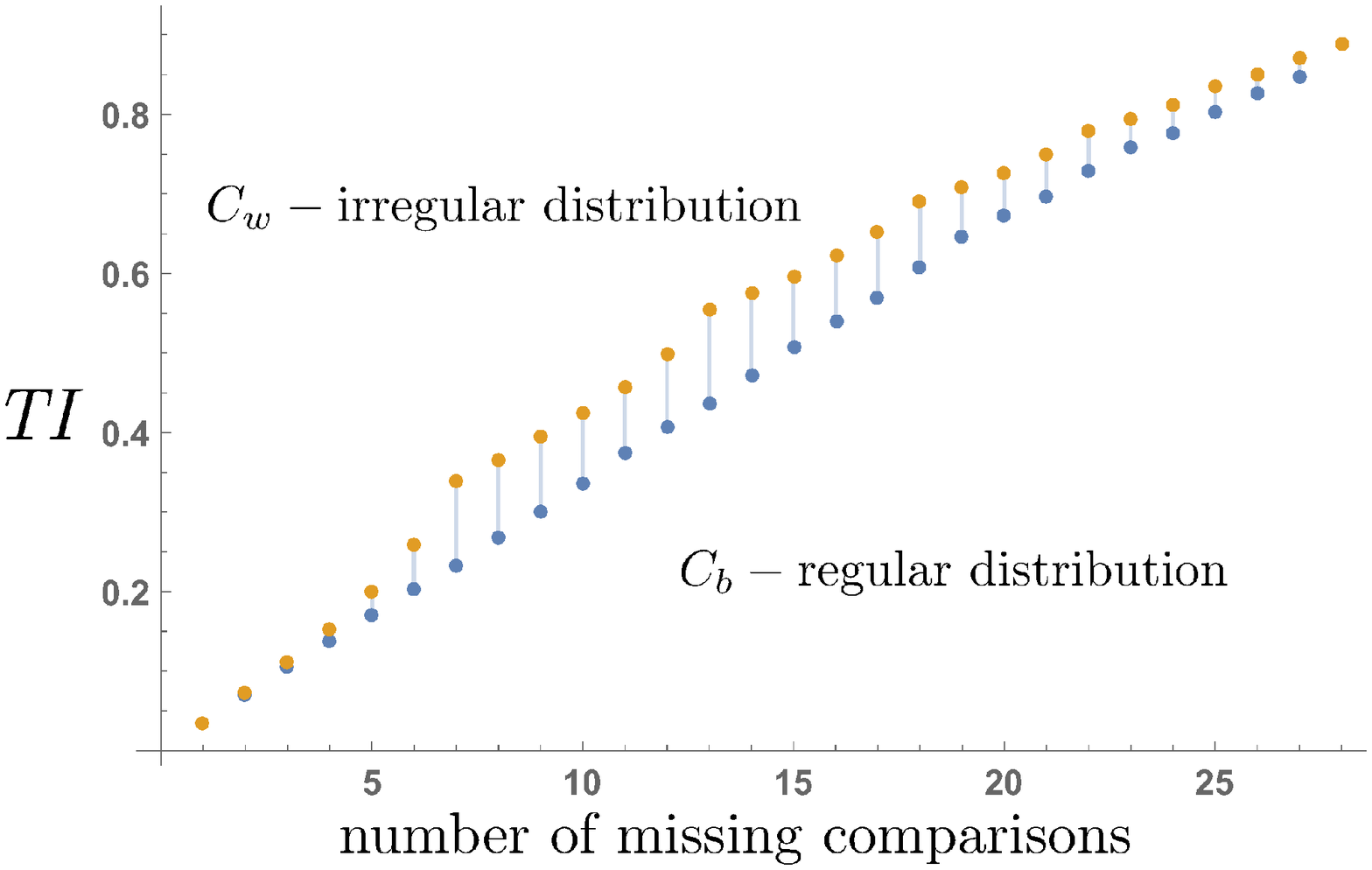}
\par\end{centering}
\label{fig:impact-tree-index}}
\par\end{centering}
\begin{centering}
\subfloat[$\alpha,\beta$-index (with $\alpha=1.5$ and $\beta=2$)]{\begin{centering}
\includegraphics[scale=0.3]{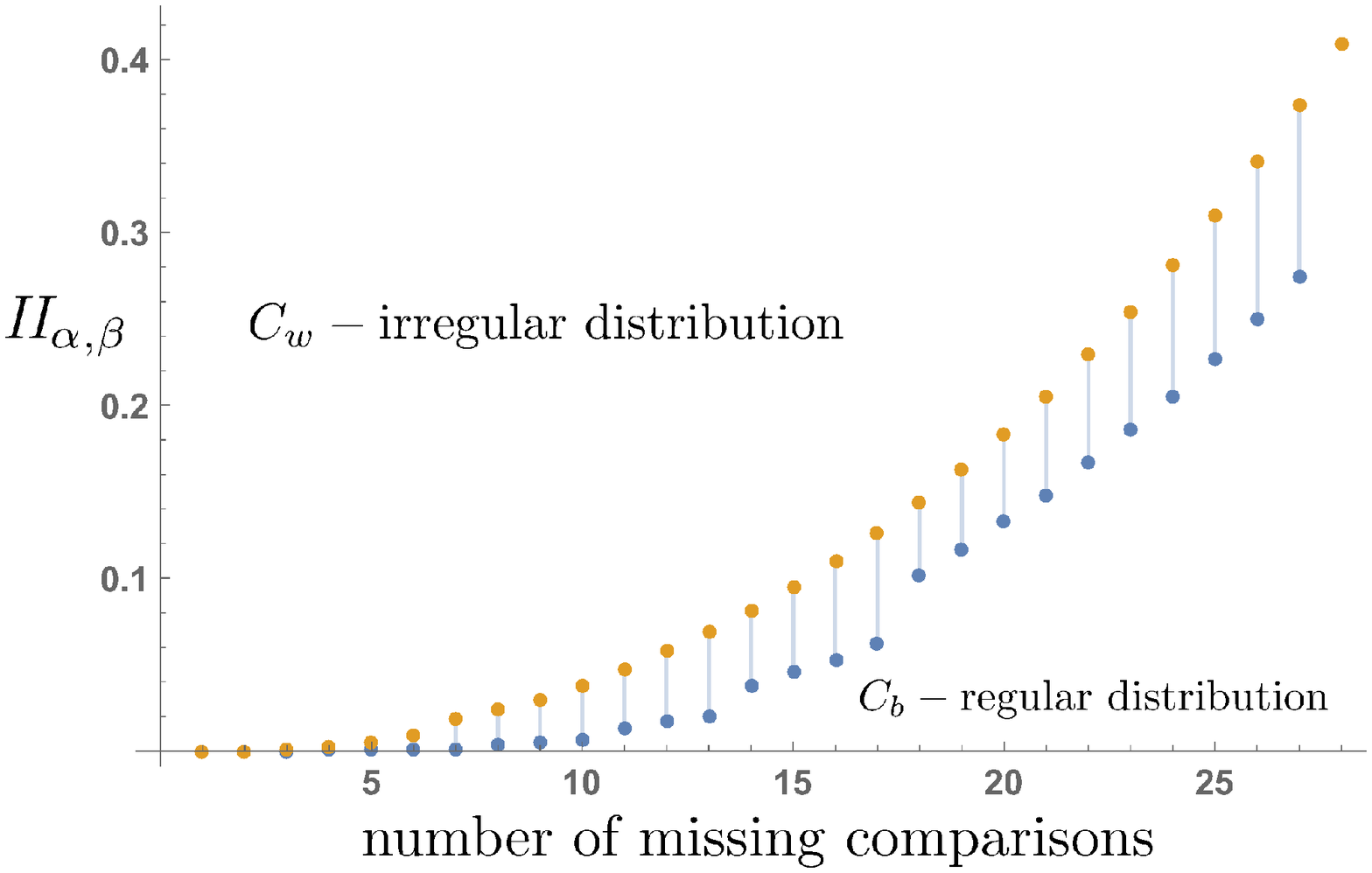}
\par\end{centering}
\label{fig:impact-alpha-beta-index}}
\par\end{centering}
\caption{Impact of the distribution of missing comparisons to indices of incompleteness,
measured in the group of random PC matrices $9\times9$ with the average
inconsistency $\textit{CI}\approx0.1$.}

\label{fig:impact-all}
\end{figure}

In the similar way we tested all the indices. In the Figure \ref{fig:impact-all}
we can see plots of $\textit{II}_{\alpha}$, $\textit{II}_{\beta}$,
$\textit{TI}$ and $\textit{II}_{\alpha,\beta}$ correspondingly. 

Although all the indices rise along the increase of the number of
missing values their increase differs from plots of sensitivity. For
a not very high number of missing values (here $14$ which is $50\%$
of all comparisons possible to remove) all the indices seem to mimic
the sensitivity charts (Fig. \ref{fig:impact-indices}). However,
for the larger numbers of missing comparisons the differences between
PC matrices formed according to $C_{b}$ and $C_{w}$ are important.
The use of $\textit{II}_{\beta}$ or $\textit{II}_{\alpha,\beta}$
can help in this case. 

\subsection{Discussion}

The first experiment (Section \ref{subsec:Relationship-between-incompleten})
clearly shows that both: inconsistency and incompleteness almost equally
contribute to the sensitivity of the given PC matrix. This means that
when assessing the quality of the matrix its completeness cannot be
ignored. On the other hand, Figures \ref{fig:SensitivityManhattanFig}
and \ref{fig:SensitivityKendallFig} suggest that when the number
of missing elements is small, the impact of this deficiency on the
final ranking is almost negligible. However, when a lot of comparisons
are missing the ranking can be significantly changed due to incompleteness. 

Since the proposed indices aim to determine not only a simple number
of missing comparisons but also their distributions in way that allows
the user to discover potential risks of vulnerability to disturbances.
In the second experiment (Section \ref{subsec:Impact-of-the}) we
analyze the influence of the distribution of missing elements to the
sensitivity of the PC method and the values of incompleteness indices. 

The experiments carried out show that all the indices grow (or at
least do not decrease) as the number of missing values increases.
Similarly all the indices get the greater values when the distribution
of missing values is potentially less favorable. However, despite
many similarities the values of indices and the values of sensitivity
are not identical. Thus, computing and analyzing the incompleteness
indices can not replace the classical sensitivity analysis. Therefore,
incompleteness indices should be treated as kind of a yardstick which
allows to quickly detect that incompleteness can be a problem and
should be improved. The great advantage of incompleteness indices
is the ease of their calculation. As all of them use the number of
missing comparisons on their inputs for $n\times n$ PC matrix we
need at most $O(n^{2})$ operations. Performing the sensitivity analysis
usually is much more time and resource consuming. Even worse, as the
sensitivity analysis tries to answer the questions how the changes
in the input data translate to the method outcome, it might happen
that the incompleteness as an actual source of problems can be overlooked.
The indices of incompleteness eliminate danger. Due to their simplicity
they are great for quick and simple test of completeness of the paired
decision data. 

\section{Summary\label{sec:Summary}}

This paper has developed four incompleteness indices for using with
the quantitative pairwise comparisons method with incomplete set of
comparisons. These indices can be used as fast and computationally
simple data quality tests. The constructed indices have been tested
in Montecarlo experiments. Carried trials showed a significant impact
of incompleteness expressed by these indices to the sensitivity of
the pairwise comparisons based decision model. Although it is clear
that the incompleteness only is just one of the factors affecting
sensitivity, the defined indices can help the decision makers to discover
the risks to sensitivity having their source in the data incompleteness. 

\section*{Acknowledgment}

The research is supported by The National Science Centre, Poland,
project no. 2017/25/B/HS4/01617 and by the Polish Ministry of Science
and Higher Education. 

\bibliographystyle{IEEEtranS}
\bibliography{papers_biblio_reviewed}

\end{document}